\documentclass[runningheads]{llncs}
\usepackage[T1]{fontenc}
\usepackage{booktabs}
\usepackage{amsmath}
\usepackage{amssymb}
\usepackage{tabularx}
\usepackage{multirow}
\usepackage{graphicx}
\usepackage{hyperref}
\usepackage{color}
\usepackage{comment}

\begin{document}

\title{ReproScore: Separating Readiness from Outcome in Research Software Reproducibility Assessment}

\titlerunning{ReproScore: Separating Readiness from Outcome}
\author{Sheeba Samuel\inst{1}\orcidID{0000-0002-7981-8504} \and
Daniel Mietchen\inst{2}\orcidID{0000-0001-9488-1870} \and
Jungsan Kim\inst{1}\and
Waqas Ahmed\inst{3}\orcidID{0000-0002-9354-3527} \and
Martin Gaedke\inst{1}\orcidID{0000-0002-6729-2912}
}
\authorrunning{Samuel et al.}
\institute{Chemnitz University of Technology \email{{sheeba.samuel, martin.gaedke}@informatik.tu-chemnitz.de} \and
FIZ Karlsruhe — Leibniz Institute for Information Infrastructure, Germany \email{daniel.mietchen@fiz-karlsruhe.de} \and
Friedrich Schiller University, Jena, Germany \email{ahmed.waqas@uni-jena.de}
}
\maketitle
\begin{abstract}
Digital libraries curate millions of research software artefacts yet lack scalable infrastructure for assessing whether those artefacts remain executable. Existing automated assessment tools treat
  static repository completeness---what a repository \emph{contains}---as a proxy for execution success---whether it \emph{runs}. We term this the \emph{readiness--outcome conflation} 
  and present \textbf{ReproScore}, a two-tier framework that explicitly separates reproducibility readiness (RRS) from reproducibility outcome (ROS), combining them into a
  coverage-adaptive Composite Score (RCS). RRS comprises 26 sub-metrics across five categories; ROS provides execution-based probes when sandbox infrastructure is available; a         
  community rubric externalises weighting priorities as versioned YAML profiles. Evaluated on 423 GitHub repositories from a large-scale ground-truth corpus spanning five failure
  modes, two complementary findings emerge: the environment category strongly discriminates failure mode, confirming static signals capture meaningful structural differences; yet RRS
  exhibits near-zero binary success correlation, empirically quantifying the readiness--outcome gap at repository scale. Together, these findings validate the architectural separation as both
  necessary and non-trivial, positioning ReproScore as scalable infrastructure for reproducibility-aware curation in digital library workflows.

\end{abstract}
\section{Introduction}
Digital libraries and research repositories---Zenodo, Software Heritage,
Figshare, and institutional data archives---now store research software
at a scale that challenges manual curation.
The practical problem facing data librarians is concrete: when a deposit
arrives, how can they efficiently assess whether it is executable, identify
quality deficiencies, and prioritise remediation?
This challenge sits within the Ingest and Archival Storage entities of
the OAIS Reference Model~\cite{lavoie2014oais} and the appraisal stage
of the DCC Curation Lifecycle~\cite{higgins2008the}---yet neither
framework supplies computable metrics for research software executability.

The scale of the underlying reproducibility problem is well documented:
fewer than one in four Jupyter notebooks can be re-executed without
error~\cite{pimentel2019a}, only 26\% of computational articles
in \emph{Science} were reproducible under varying
effort~\cite{stodden2018empirical}, and 90\% of surveyed researchers
acknowledged a reproducibility crisis~\cite{baker20161500}.
At repository intake scale, these rates imply that the majority of
deposits labelled as computational research cannot be re-executed by a
curator who tries. This is a library infrastructure problem, not merely a
research culture problem.

Existing automated tools address this gap incompletely and from opposing
directions.
Static analysis tools---SciScore~\cite{bandrowski2022sciscore},
SOMEF~\cite{mao2019somef}, and README-based
classifiers~\cite{akdeniz2023end}---assess documentation completeness
and metadata coverage: what a deposit \emph{contains}.
Execution-based approaches---CODECHECK~\cite{nust2021codecheck},
Whole Tale~\cite{brinckman2019computing}, and
repo2docker~\cite{forde2018reproducible}---assess whether code
\emph{runs}, but require expert human involvement or sandboxed
infrastructure and cannot operate at collection intake scale.
Automated notebook analysis~\cite{pimentel2019a,samuel2024computational}
has characterised execution failure patterns in large corpora but stops
short of providing a scoring framework for curatorial decision-making.
The critical gap is that \emph{static artefact quality and execution
outcome are genuinely different quantities}: a repository with a pinned
\texttt{requirements.txt}, a Zenodo data pointer, and a structured README
is an excellent static deposit that may nonetheless fail at install time
due to issues like a transitive dependency conflict---a failure no static tool can
observe.

We call this the \emph{readiness--outcome conflation} and present
\textbf{ReproScore}, a repository curation framework that addresses it
within digital library workflows.
ReproScore functions as a \emph{failure-mode characterisation}
instrument: it tells a curator \emph{what kind} of reproducibility
problem a repository has (missing environment specification, inaccessible
data, portability defects, absent determinism signals), not merely
whether it will succeed.
This diagnostic framing, combined with a transparent rubric governance
mechanism, constitutes the primary contribution.

\textbf{Contributions.}
This paper makes the following contributions:
(1) A \textbf{conceptual and computational separation} between readiness (static, available at intake) and execution outcome (optional), formalised as a two-tier architecture (RRS/ROS/RCS) that
addresses the readiness--outcome conflation in digital library curation workflows.
RRS is the primary empirical contribution; ROS and RCS constitute a partially validated architectural extension.
(2) A \textbf{five-category, 26-sub-metric readiness model} (RRS) with expert-informed weights, a non-linear gate encoding curation policy, and file-level evidence provenance.
(3) A \textbf{community rubric mechanism} externalising assessment priorities as a versioned YAML profile.
(4) An \textbf{empirical evaluation on 423 GitHub repositories} from a large-scale ground-truth corpus~\cite{samuel2024computational} showing static readiness strongly discriminates failure mode
($H = 96.89$, $p < 0.001$).

The remainder of this paper is structured as follows.
Section~\ref{sec:related} discusses related work.
Section~\ref{sec:model} details the scoring architecture and rubric mechanism.
Section~\ref{sec:evaluation} presents the empirical evaluation.
Section~\ref{sec:discussion} and Section~\ref{sec:conclusion} provide discussion and concluding remarks.

% ============================================================
\section{Background and Related Work}
\label{sec:related}
% ============================================================

\textbf{Reproducibility in digital library curation.}
The OAIS Reference Model~\cite{lavoie2014open} identifies Ingest and
Archival Storage as functional entities where quality assessment occurs,
but specifies no executable quality metrics for research software.
The DCC Curation Lifecycle~\cite{higgins2008the} frames quality appraisal
as a continuous obligation across the deposit lifecycle.
Software Heritage~\cite{di2017software} preserves source code at
collection scale; Zenodo, Figshare, and institutional repositories
accept heterogeneous deposits including Jupyter notebooks, R Markdown
scripts, and containerised workflows.
None of these platforms provides systematic, automated assessment of
executability at intake---the gap ReproScore targets.

\noindent\textbf{FAIR for software.}
The FAIR Guiding Principles~\cite{wilkinson2016fair} and FAIR4RS~\cite{chue2022fair}
articulate reusability desiderata but supply no computable executability scores.
Tools such as \texttt{howfairis}~\cite{spaaks2021howfairis},
FAIRsFAIR~\cite{devaraju2020fairsfair}, and
FAIR-Checker~\cite{gaignard2023fair} assess FAIR compliance at the metadata level.
CodeMeta~\cite{jones2017codemeta} and RO-Crate~\cite{soiland2022packaging}
provide description vocabularies; neither links metadata to execution potential.

\noindent\textbf{Static quality analysis.}
SciScore~\cite{bandrowski2022sciscore} analyses biomedical paper methods
sections for reporting completeness, operating on paper text rather than
code.
Better Code Hub and SonarQube~\cite{campbell2013sonarqube} assess
maintainability and security without targeting reproducibility.
Akdeniz et al.~\cite{akdeniz2023end} parse README files through
transformer models to generate reproducibility scores from section
completeness---structurally identical to the conflation they aim to
address.
Unlike general-purpose software quality models (ISO/IEC 25010; CHAOSS \cite{goggins2021open}), ReproScore operationalises executability-specific criteria essential for reproducibility assessment.

\noindent\textbf{Execution-based evaluation.}
CODECHECK~\cite{nust2021codecheck} provides an expert peer-review
workflow for individual submissions.
Trisovic et al.~\cite{trisovic2022large} execute Harvard Dataverse
R scripts at scale, finding version conflicts as the dominant failure
mode---motivating ReproScore's emphasis on environment specification.
Costa et al.~\cite{costa2024evaluating} compare eight reproduction
tools, finding that design choices critically shape execution outcomes.
Recent LLM-agent benchmarks---PaperBench~\cite{starace2025paperbench},
CORE-Bench~\cite{siegel2024core}, SUPER~\cite{bogin2024super},
REPRO-Bench~\cite{hu2025repro}---evaluate automated reproduction on
individual papers.
These execution-based agents and static readiness tools are
complementary: agents require functional code to orchestrate;
static readiness scoring provides a pre-execution triage layer that
identifies \emph{why} code is unlikely to function before any
execution resource is committed---and reduces the environmental cost
of curation by screening out unexecutable deposits before
compute-intensive container builds and sandbox runs are attempted.
A consistent pattern is that static documentation quality and
execution success are systematically conflated~\cite{carlson2025nerve,pimentel2019a,trisovic2022large,vangala2025ai}.
ReproScore addresses this directly.

% ============================================================
\section{The ReproScore Scoring Architecture}
\label{sec:model}
\begin{figure}[ht]
\centering
\includegraphics[width=\textwidth]{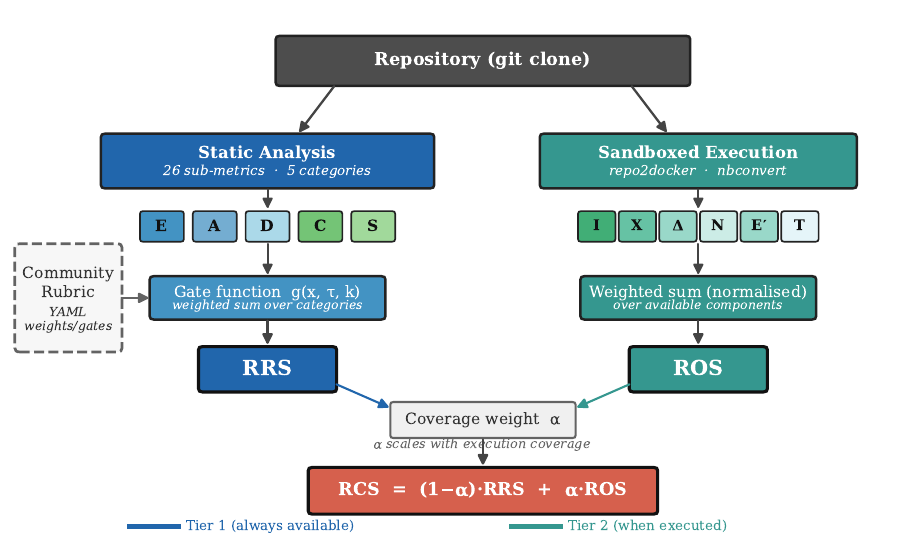}
\caption{ReproScore two-tier architecture. Tier~1 (static analysis,
  always available at intake) computes RRS (cf.\ Section~\ref{sec:rrs}) from 26~sub-metrics (cf.\ Table~\ref{tab:submetrics}) in 5 categories  (cf.\ Table~\ref{tab:rrs-categories}) via a
  community-configurable rubric (cf. Section~\ref{sec:rubric}). Tier~2 (sandboxed execution, optional)
  computes ROS (cf.\ Section~\ref{sec:ros}) from up to six execution probes. RCS  (cf.\ Section~\ref{sec:rcs}) blends both tiers
  via a coverage weight~$\alpha$ that scales with collected execution
  evidence, allowing the composite metric to degrade gracefully from
  RCS = RRS (no execution) to RCS $\approx$ ROS (full execution coverage).}
\label{fig:architecture}
\end{figure}

% ============================================================
Figure~\ref{fig:architecture} illustrates the ReproScore two-tier pipeline.

% ============================================================
\subsection{Reproducibility Readiness Score (RRS)}
\label{sec:rrs}
% ============================================================

\subsubsection{Category structure and weights.}

RRS is computed over five top-level categories, each capturing a distinct
dimension of reproducibility readiness.
Table~\ref{tab:rrs-categories} lists the categories, default weights,
gate parameters, and sub-metric counts.
\begin{table}[ht]
\centering
\caption{RRS categories, default weights, gate parameters, and sub-metric
  counts. Weights sum to 1.0. 
  Core categories (E, A) use steeper gates;
  quality categories (D, C, S) use lenient gates.}
\label{tab:rrs-categories}
\begin{tabular}{@{}l|l|l|l|l|l@{}}
\toprule
\textbf{Symbol} & \textbf{Category and guiding question}
  & \textbf{Weight $w_i$}
  & $\boldsymbol{\tau_i}$ & $\boldsymbol{k_i}$ & \textbf{Sub-metrics} \\
\midrule
$E$ & \begin{tabular}[t]{@{}l@{}}{\bf E}nvironment specification\\
      \textit{\small Can the computational environment be reconstructed?}
      \end{tabular}
    & 0.30 & 40 & 1.5 & 4 \\[6pt]
$A$ & \begin{tabular}[t]{@{}l@{}}Data {\bf a}ccessibility\\
      \textit{\small Can the required data be obtained and used?}
      \end{tabular}
    & 0.25 & 30 & 1.5 & 4 \\[6pt]
$D$ & \begin{tabular}[t]{@{}l@{}}{\bf D}ocumentation\\
      \textit{\small Can a third party understand how to execute this?}
      \end{tabular}
    & 0.20 & 20 & 1.2 & 7 \\[6pt]
$C$ & \begin{tabular}[t]{@{}l@{}}{\bf C}ode portability\\
      \textit{\small Will the code run on a machine other than the author's?}
      \end{tabular}
    & 0.15 & 25 & 1.2 & 4 \\[6pt]
$S$ & \begin{tabular}[t]{@{}l@{}}Reproducibility {\bf s}ignals\\
      \textit{\small Will repeated executions produce the same verifiable result?}
      \end{tabular}
    & 0.10 & 30 & 1.2 & 7 \\
\midrule
    & \textit{Total} & 1.00
    & \multicolumn{3}{l}{26 atomic sub-metrics} \\
\bottomrule
\end{tabular}
\end{table}
Both inter-category weights $w_i$ and within-category weights $w_j$ are expert-informed default choices, not data-optimised parameters, and exposed through the community rubric (Section~\ref{sec:rubric}).
Environment specification receives $w_E = 0.30$ because missing or conflicting dependencies are the dominant execution failure mode~\cite{trisovic2022large,samuel2024computational}.
Data accessibility receives $w_A = 0.25$ because inaccessible or unresolvable input data is the second most common barrier to re-execution~\cite{stodden2018empirical}: a pipeline that cannot obtain its inputs cannot be executed regardless of environment quality. Documentation receives $w_D = 0.20$ because the absence of execution-relevant instructions---entry points, install steps, expected
outputs---forces a third party to reverse-engineer the workflow 
before any execution attempt can be made, compounding other failure modes. Code portability receives $w_C = 0.15$ because source-code properties that cause failure when code is moved across machines---absolute paths, undeclared imports, hardcoded credentials---are detectable statically and directly predictive of execution failure independent of environment quality~\cite{trisovic2022large}.
Reproducibility signals receives $w_S = 0.10$ because workflow practices and determinism commitments---seed management, execution order, output declarations---are necessary for result verification but
presuppose that execution itself succeeds; they are therefore downstream of the four primary categories.

\subsubsection{Gate function.}
A power-law gate function penalises sub-threshold failures
non-linearly: $E = 2$ (no specification) should not score comparably
to $E = 35$ (partial specification) under a linear scale.
The gate encodes a curation policy claim rather than a
performance mechanism.
ReproScore defines $g : [0,100] \times \mathbb{R}_{>0} \times
\mathbb{R}_{>1} \to [0,1]$ as:

\begin{equation}
  g(x,\,\tau,\,k) \;=\;
  \begin{cases}
    \dfrac{x}{100} & \text{if } x \geq \tau \\[10pt]
    \left(\dfrac{x}{\tau}\right)^{\!k} \cdot \dfrac{\tau}{100}
      & \text{if } x < \tau
  \end{cases}
  \label{eq:gate}
\end{equation}

\noindent where $x \in [0,100]$ is the raw sub-score, $\tau$ is the
threshold below which compression begins, and $k > 1$ controls
steepness.
Above $\tau$, $g$ is linear.
Below $\tau$, $g$ applies a power-law contraction: at $x = \tau/2$,
the contribution is $2^{-k} \cdot \tau/100$---less than half the
linear value.
Core categories use $k = 1.5$ (steep); quality categories use $k = 1.2$
(lenient).
Robustness analysis (Section~\ref{sec:robustness}) shows the gate has
negligible predictive impact (AUC range $\approx 0.008$);
its value is normative.

\subsubsection{RRS formula.}

The Reproducibility Readiness Score is computed as:

\begin{equation}
  \mathrm{RRS} \;=\; 100 \cdot \sum_{i \in \{E,A,D,C,S\}} w_i \cdot g(x_i,\,\tau_i,\,k_i)
    \;-\; P_{\mathrm{hard}}(E, A) \;-\; P_{\mathrm{seed}}
  \label{eq:rrs}
\end{equation}

\noindent where $x_i \in [0, 100]$ is the raw category score, $w_i$ and
$(\tau_i, k_i)$ are the weights and gate parameters from
Table~\ref{tab:rrs-categories}, and the penalty terms are:

\begin{align}
  P_{\mathrm{hard}}(E, A) &\;=\;
    20 \cdot \mathbf{1}[E < 10] \;+\; 15 \cdot \mathbf{1}[A < 10]
  \label{eq:p-hard} \\[4pt]
  P_{\mathrm{seed}} &\;=\;
    10 \cdot \mathbf{1}[\sigma < 50]
  \label{eq:p-seed}
\end{align}

\noindent where $\sigma$ is the seed management score (sub-metric of
$S$).
$P_{\mathrm{hard}}$ penalises complete absence of environment
specification or data artefacts---conditions under which reproduction
is virtually impossible.
$P_{\mathrm{seed}}$ penalises repositories with stochastic operations
but no seed-setting calls.
Penalty magnitudes are calibrated so that each penalty approximates
the maximum weight contribution of the penalised category:
$-20$ pts for $E < 10$ corresponds to $w_E \times 100 \times 0.67$;
$-15$ pts for $A < 10$ to $w_A \times 100 \times 0.60$.
$\mathrm{RRS} \in [0, 100]$ after clamping.

\subsubsection{Sub-metric definitions}
Table~\ref{tab:submetrics} presents the full 26-sub-metric taxonomy.
Each sub-metric is computed by deterministic, rule-based analysis of
the locally cloned repository---no execution required.
Sub-metrics are grounded in execution failure patterns from large-scale
reproduction studies~\cite{pimentel2019a,samuel2024computational,trisovic2022large},
FAIR4RS criteria~\cite{chue2022fair}, and static signals from the FAIR
Jupyter knowledge graph~\cite{samuel2024FAIR} (e.g.\
\texttt{notebook\_exec\_order}, \texttt{markdown\_code\_ratio}).

\begin{table}[!tbp]
\centering
\footnotesize
\caption{ReproScore 26 atomic sub-metrics with within-category weights $w_j$
  (sum to 1.0 within each category).
  Type: Bin = binary (0/100); Cont = continuous; Tier = tiered discrete levels.
  Full measurement heuristics are available in the source repository.
  }
\label{tab:submetrics}
\begin{tabularx}{\linewidth}{@{}l|l|r|X|l|@{}}
\toprule
\textbf{Cat.} & \textbf{Sub-metric} & $w_j$ & \textbf{Rationale} & \textbf{Type} \\
\midrule
\multirow{4}{*}{$E$}
  & Dep.\ pinning        & 0.25 & Lockfile $>$ exact pins $>$ partial $>$ absent; tiered by reproducibility guarantee & Tier \\
  & Container spec       & 0.30 & Container file presence and quality; pinned base + RUN is strongest & Tier \\
  & Env.\ bootstrap      & 0.25 & One-command environment creation (install script, Makefile target) & Bin \\
  & Runtime version      & 0.20 & Explicit Python/R version declaration (.python-version, runtime.txt, etc.) & Bin \\
\midrule
\multirow{4}{*}{$A$}
  & Data description     & 0.20 & Dedicated data documentation; tiered by word count / section richness & Tier \\
  & Data pointer         & 0.30 & Tiered by archival permanence: DOI $>$ institutional $>$ platform URL $>$ local file & Tier \\
  & Workflow orch.       & 0.20 & End-to-end workflow tool (Snakemake, DVC, Nextflow) $>$ Makefile $>$ pipeline script & Tier \\
  & Data acquisition     & 0.30 & Automated download present (DVC tracking, wget/curl/API calls to data archives) & Bin \\
\midrule
\multirow{7}{*}{$D$}
  & Doc.\ structure      & 0.25 & Fraction of 4 execution-relevant README sections (install, run, expected output, requirements) & Cont \\
  & Install instructions & 0.20 & Completeness of install guidance; one-command $>$ multi-step $>$ vague & Tier \\
  & Usage examples       & 0.20 & Runnable command in code fence $>$ any code block $>$ examples directory & Tier \\
  & Inline explanation   & 0.15 & Notebook md/code ratio and script comment density; both target $\geq 0.5$ / $\geq 20\%$ & Cont \\
  & Entry point          & 0.10 & Clear first command to execute (run.sh, main.py, Makefile run target) & Bin \\
  & Docstring coverage   & 0.05 & Inline docstrings on public functions/classes & Cont \\
  & Reuse metadata       & 0.05 & LICENSE + CITATION.cff + codemeta.json; tiered by count (transparency, not execution) & Tier \\
\midrule
\multirow{4}{*}{$C$}
  & No absolute paths    & 0.40 & Machine-specific paths (/home/user/, C:\textbackslash{}Users\textbackslash{}) absent from source files \& notebooks & Cont \\
  & Import resolvability & 0.35 & Third-party imports cross-referenced against declared dependencies & Cont \\
  & No hardcoded creds   & 0.15 & Credential assignment patterns (API keys, tokens) absent from source & Cont \\
  & No silent failures   & 0.10 & Bare \texttt{except:pass} (hides execution errors) absent from source & Cont \\
\midrule
\multirow{7}{*}{$S$}
  & Seed management      & 0.30 & $|\mathcal{F}_\text{seed}|/|\mathcal{F}_\text{rand}|$: stochastic files with seed-setting calls & Cont \\
  & Notebook exec.\ order & 0.20 & Notebooks have monotonically increasing execution counts (run top-to-bottom) & Cont \\
  & Test file presence   & 0.18 & $\min(|\mathcal{T}|/2,1)$; test suite existence & Cont \\
  & Expected outputs     & 0.12 & Reference output directory or committed figures present & Tier \\
  & CI presence          & 0.10 & Any CI configuration file present & Bin \\
  & Config externalised  & 0.06 & Experimental parameters in config files or CLI args (not hardcoded) & Tier \\
  & Hardware requirements & 0.04 & If GPU packages detected: CUDA/hardware requirements declared; otherwise N/A & Bin \\
\bottomrule
\end{tabularx}
\end{table}

% ============================================================
\subsection{Reproducibility Outcome Score (ROS)}
\label{sec:ros}
% ============================================================
When a repository is executed in a sandboxed environment, ReproScore
computes an execution-based outcome score comprising six components:
install success $I$ (0.30), execution success $X$ (0.25), output
determinism $\Delta$ (0.20), notebook execution rate $N$ (0.10), import
success rate $E'$ (0.10, directly measuring dependency resolution), and
test pass rate $T$ (0.05).
All components are optional; normalisation adjusts for missing components
via a weighted sum over available probes only:
\begin{equation}
  \mathrm{ROS} \;=\;
  \frac{\displaystyle\sum_{j} v_j \cdot y_j \cdot
    \mathbf{1}[y_j\text{ available}]}
       {\displaystyle\sum_{j} v_j \cdot
    \mathbf{1}[y_j\text{ available}]}
  \label{eq:ros}
\end{equation}

\noindent where $y_j \in [0, 100]$ is the score for ROS component $j$
and $v_j$ is its weight.
When no components are available, ROS is undefined ($\perp$) and the
system falls back to $\mathrm{RCS} = \mathrm{RRS}$.

$I$ and $E'$ are \emph{causally independent}: $I$ tests
package-manager resolution; $E'$ tests runtime import success---both
can diverge (confirmed empirically in Section~\ref{sec:ros-validation}).
$X$ is binary (overall outcome); $N$ is continuous (fraction of
notebooks completing without error).
$\Delta$ compares output cells across two sandboxed runs via nbdime
(exact match for non-numeric; tolerance $10^{-6}$ for numeric).

% ============================================================
\subsection{Composite Score and Coverage Weighting (RCS)}
\label{sec:rcs}
% ============================================================

The coverage weight $\alpha$ measures what fraction of maximum
execution evidence has been collected, capped at $\alpha_{\max} = 0.70$:

\begin{equation}
  \alpha \;=\; \min\!\left(\sum_{j} v_j \cdot
    \mathbf{1}[y_j\text{ available}],\;\; 1 \right) \cdot \alpha_{\max}
  \label{eq:alpha}
\end{equation}

\noindent with floor $\alpha \geq \alpha_{\min} = 0.10$ whenever any
ROS component is available.
The ceiling $\alpha_{\max} = 0.70$ encodes the principle that static
readiness is never entirely dominated by execution outcomes: even
perfect execution does not measure data documentation, README
completeness, or portability.
The composite score is:

\begin{equation}
  \mathrm{RCS} \;=\;
  \begin{cases}
    \mathrm{RRS}
      & \text{if } \mathrm{ROS} = \perp \\[4pt]
    (1 - \alpha) \cdot \mathrm{RRS} \;+\; \alpha \cdot \mathrm{ROS}
      & \text{otherwise}
  \end{cases}
  \label{eq:rcs}
\end{equation}

For partial empirical validation of the RCS blending formula, see  Section~\ref{sec:ros-validation}.

% ============================================================
\subsection{Community Rubric Mechanism}
\label{sec:rubric}
% ============================================================
Assessment priorities differ across communities.
A bioinformatics data archive may weigh data accessibility more
heavily than CI configuration; a software-centric repository may
prioritise portability.
ReproScore externalises these priorities through a
\emph{community rubric}: a named, versioned YAML file overriding
default category weights $w_i$ and gate thresholds.
The \texttt{RubricEngine} validates $\sum w_i = 1.0 \pm 0.01$ and
recomputes the score using the community-specified profile.
An illustrative FAIR-aligned bioinformatics override could look as follows:

\begin{small}
\begin{verbatim}
name: bioinformatics-v1
version: "1.0"
categories:
  E: {weight: 0.35, tau: 40, k: 1.5}
  A: {weight: 0.40, tau: 30, k: 1.5}  # FAIR data priority
  D: {weight: 0.10, tau: 20, k: 1.2}
  C: {weight: 0.05, tau: 25, k: 1.2}
  S: {weight: 0.10, tau: 30, k: 1.2}
\end{verbatim}
\end{small}

A platform assigning $w_A = 0.40$ expresses a versioned, auditable
curation policy, 
in contrast to scoring heuristics embedded opaquely in source code.
Rubric files are portable across institutions: archives can publish
their weighting rationale as a citable artefact, enabling policy
comparison and reuse.

\medskip\noindent\textbf{Rubric governance and federation.}
Community rubric files function as governance artefacts, not merely
configuration.
An institution publishing its rubric declares a curation policy
independently citable by peer archives, enabling cross-institutional
score comparability wherever the same rubric version was applied.
Archives can federate: a consortium of biomedical repositories might
adopt a shared baseline rubric, then document domain-specific overrides
with rationale.
When thresholds or weights are revised, affected scores can be
recomputed from stored provenance records without re-cloning, and the
version history makes policy evolution auditable.
The rubric schema is designed for serialisation as a CodeMeta \cite{jones2017codemeta} property
or RO-Crate \cite{soiland2022packaging} contextual entity, enabling score-with-policy packaging
within standard research object formats.

% ============================================================
\section{Evaluation}
\label{sec:evaluation}
% ============================================================
\subsection{Corpus and Protocol}
%The evaluation builds on a large-scale computational reproducibility corpus~\cite{samuel2024computational} comprising \textbf{27,271~Jupyter notebooks} from \textbf{2,660~GitHub repositories} associated with \textbf{3,467~PubMed Central publications}.
We evaluate the RRS scoring model on a stratified sample of Python repositories from a large-scale execution ground-truth corpus~\cite{samuel2023dataset,samuel2024computational}, which records the Python traceback of each notebook's first failing cell (or NULL for success).
Repository-level failure modes aggregate notebook records via priority ordering (install errors $\succ$ import errors $\succ$ file-not-found errors $\succ$ code errors $\succ$ success); a repository is labelled \emph{success} only when all its notebooks yield NULL.
We stratify by failure mode, targeting \textbf{86 repositories per class}:
\emph{success} (all execution failure reasons $=$ NULL),
\emph{install\_dep} (\texttt{<Install Dependency Error>}),
\emph{missing\_module} (\texttt{ModuleNotFoundError}/\texttt{ImportError}), \mbox{}
\emph{missing\_data}  (\texttt{FileNotFoundError}/ 
network errors),
and \emph{code\_error} (\texttt{TypeError}/\texttt{NameError}/\texttt{SyntaxError}
and related runtime exceptions).
Of 430 sampled repositories, 423 were scored (7 excluded due to clone or
computation errors), yielding 84--85 per class.
Each repository was shallow-cloned and scored by static RRS analysis
only; no execution was performed.
Per-repository provenance JSON and cloned SHA are archived with the
dataset~\cite{samuel2026reproscore}.
The sample is restricted to Python Jupyter notebook repositories to match the execution environment of the original dataset.
The five-class failure mode is the primary ground truth for
failure-mode discriminability (Kruskal-Wallis $H$, pairwise Cohen's
$d$); binary success/failure is a secondary non-collapse check only.
All statistical tests 
were computed using SciPy~\cite{virtanen2020scipy} (Python~3).

\subsection{Failure-Mode Characterisation}
\label{sec:characterisation}
\noindent\textbf{Evaluation goals.}
The primary criterion is whether each category structurally
separates repositories by failure type---a curator needs to know
\emph{what kind} of problem a repository has, not its failure
probability.
AUC-ROC is used solely as a non-collapse probe in
Section~\ref{sec:robustness}.
Table~\ref{tab:ablation-validity} reports category-level statistics.

\begin{table}[!ht]
\centering
\caption{Category-level discriminability on the 423-repository sample.
  $H$: Kruskal-Wallis statistic, $df = 4$ (primary criterion).
  $r_{pb}$: point-biserial with binary success label (secondary;
  near-zero is expected for a failure-mode characterisation instrument).}
\label{tab:ablation-validity}
\begin{tabular}{@{}llrrrr@{}}
\toprule
\textbf{Cat.} & \textbf{Name} & $w_i$ & $r_{pb}$ & $p_{pb}$
  & $H$ ($p_{KW}$) \\
\midrule
$E$ & Environment spec.  & 0.30 & $-0.014$ & 0.767 & $96.89\;(<\!0.001)$ \\
$A$ & Data accessibility & 0.25 & $+0.016$ & 0.745 & $16.12\;(0.003)$ \\
$D$ & Documentation      & 0.20 & $+0.028$ & 0.572 & $53.09\;(<\!0.001)$ \\
$C$ & Code portability   & 0.15 & $+0.060$ & 0.215 & $56.55\;(<\!0.001)$ \\
$S$ & Repro.\ signals    & 0.10 & $+0.153^{**}$ & 0.002 & $40.19\;(<\!0.001)$ \\
\bottomrule
\end{tabular}
\end{table}

\begin{figure}[t]
\centering
\includegraphics[width=\textwidth]{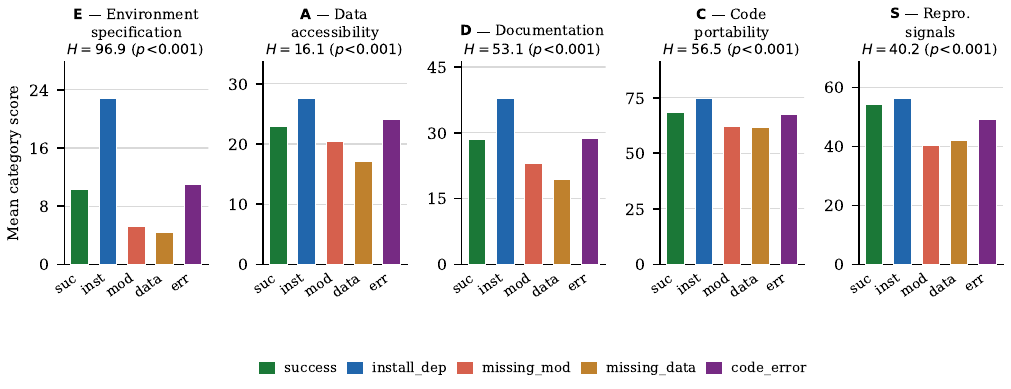}
\caption{Mean category score by failure mode (423 repositories; 84--85
  per class). The $E$ panel illustrates the detection paradox:
  \emph{install\_dep} scores highest on environment specification yet
  fails at install time; \emph{success} scores lower.
  Kruskal-Wallis $H$ and $p$-values per panel.}
\label{fig:profiles}
\end{figure}

\medskip\noindent\textbf{Finding 1: $E$ discriminates failure mode,
not failure probability.}
The environment category achieves the highest Kruskal-Wallis
$H$ ($H = 96.89$, $p < 0.001$) yet near-zero binary correlation
($r_{pb} = -0.014$, $p = 0.767$).
The mechanism is an \emph{E detection paradox}: \emph{install\_dep}
repositories score highest on $E$ (mean 22.8)---they specify
environments explicitly, but with version conflicts---while
\emph{missing\_module} (5.2) and \emph{missing\_data} (4.4) score
near zero.
Critically, \emph{success} repositories score lower ($E = 10.3$) than
\emph{install\_dep}, so directional cancellation across the binary
label produces near-zero (slightly negative) correlation.
Figure~\ref{fig:profiles} makes this concrete.
Pairwise effect sizes confirm failure-mode separability: $E$ separates
\emph{install\_dep} from \emph{missing\_module} with Cohen's $d = 1.16$
($p < 0.001$, KS $= 0.68$) and from \emph{missing\_data} with
$d = 1.25$ ($p < 0.001$, KS $= 0.68$).
For a curator, this means a low-$E$ score points to underspecification
(most likely a missing-module or missing-data failure), while a
high-$E$ score with execution failure points to a version
conflict---actionable information that binary prediction cannot provide.

\medskip\noindent\textbf{Finding 2: $C$ and $D$ discriminate failure
mode through complementary mechanisms.}
$C$ achieves the second-strongest discriminability ($H = 56.55$,
$p < 0.001$), driven by \texttt{import\_resolvability}: repositories
without dependency specification files receive score~$= 0$, correctly
penalising \emph{missing\_module} and \emph{missing\_data} repositories
(which predominantly lack dependency files), while \emph{install\_dep}
repositories (which specify dependencies, if incorrectly) score highest
($\bar{C} = 74.6$).
$D$ achieves $H = 53.09$ ($p < 0.001$), reflecting that well-documented
repositories cluster in the \emph{install\_dep} and \emph{code\_error}
modes---repositories that at least partially executed---rather than in
\emph{missing\_module} and \emph{missing\_data} modes where
documentation is sparse.

\medskip\noindent\textbf{Finding 3: Weak binary association confirms
failure-mode characterisation as the appropriate evaluation criterion.}
All $|r_{pb}| \leq 0.153$; only $S$ achieves binary significance
($p = 0.002$).
At the sub-metric level, $r_{pb}$ ranges from $-0.079$ to $+0.145$;
21 of 26 sub-metrics have $|r_{pb}| < 0.08$.
Five sub-metrics are nominally significant ($p < 0.05$):
\texttt{notebook\_exec\_order} ($r_{pb} = +0.145$, $p = 0.003$),
\texttt{seed\_management} ($+0.113$, $p = 0.020$),
\texttt{no\_absolute\_paths} ($+0.107$, $p = 0.027$),
\texttt{ci\_presence} ($+0.103$, $p = 0.035$), and
\texttt{silent\_failure\_masking} ($+0.099$, $p = 0.041$).
Applying Benjamini-Hochberg correction (FDR $= 0.05$, $m = 26$ tests),
no sub-metric survives correction (minimum $q = 0.078$ for
\texttt{notebook\_exec\_order}); all five nominally significant results
are treated as exploratory.
These weak associations are consistent with the readiness--outcome gap
in this corpus.
A post-hoc contingency analysis of $A = 0$ rates by failure mode shows
that \emph{install\_dep} repositories have the lowest $A = 0$ rate
(8\%) while \emph{missing\_data} and \emph{success} repositories have
the highest (both 24\%) ($\chi^2 = 9.89$, $df = 4$, $p = 0.042$).

\medskip\noindent\textbf{Worked example.}
A representative \emph{install\_dep} repository scores $\mathrm{RRS} = 54$
(pinned requirements, Zenodo data pointer, structured README) yet
fails at install due to a transitive version conflict; a \emph{success}
repository with $\mathrm{RRS} = 14$ (no pins, no container) executes
successfully because implicit dependencies happen to resolve. Together, these two repositories illustrate the
readiness--outcome gap.

\subsection{Robustness Diagnostics}
\label{sec:robustness}

The following diagnostics use AUC-ROC as a non-collapse probe---verifying
the composite does not implicitly behave as a binary classifier---rather
than as an optimisation target.

\noindent\textbf{Weight stability.}
Rank-stability analysis (each weight varied $\pm 50\%$ with proportional
redistribution; 20 steps per category) yields Kendall's $\tau \geq 0.911$
across all five categories, confirming stable repository ordering under
substantial weight perturbation.
A weight grid search over the simplex (126 configurations) finds maximum
AUC $= 0.581$---only $+0.045$ above the default $0.536$; the AUC-optimal
configuration concentrates weight on $S$, the category with highest binary
correlation.
The community rubric makes any such policy adjustment explicit and
auditable.

\noindent\textbf{Gate robustness.}
Sweeping $\tau$ from 10 to 70 yields AUC $0.528$--$0.536$ (span
$0.008$); linear ($k = 1$) and default ($k = 1.5$) exponents achieve
identical AUC ($0.536$).
The gate is empirically insensitive because many sub-metrics return
binary values; its significance is normative, encoding a curation
policy claim about sub-threshold failures.

\noindent\textbf{Leave-one-category-out (Table~\ref{tab:loco}).}
Removing $E$ or $A$ \emph{improves} binary AUC ($+0.028$, $+0.020$
respectively)---both categories add noise to binary discrimination
precisely because they characterise failure mode rather than predict it.
Removing $C$ or $S$ degrades AUC ($-0.022$ each), reflecting their
modest correlation with binary success.
The LOCO span is $0.050$ units; all deltas fall within the 95\%
bootstrap CI $[0.471, 0.602]$, confirming failure-mode characterisation
and binary prediction are structurally distinct properties.

\begin{table}[!ht]
\centering
\caption{Leave-one-category-out AUC. Baseline $= 0.536$
  [0.471, 0.602] (95\% CI). Positive $\Delta$: removal
  \emph{increases} binary AUC.}
\label{tab:loco}
\begin{tabular}{@{}lrr@{}}
\toprule
\textbf{Removed} & \textbf{AUC} & $\boldsymbol{\Delta}$\textbf{AUC} \\
\midrule
None (full model)          & 0.536 & --- \\
$-S$ (Repro.\ signals)     & 0.514 & $-0.022$ \\
$-C$ (Code portability)    & 0.513 & $-0.022$ \\
$-D$ (Documentation)       & 0.536 & $+0.001$ \\
$-A$ (Data accessibility)  & 0.556 & $+0.020$ \\
$-E$ (Env.\ spec.)         & 0.564 & $+0.028$ \\
\bottomrule
\end{tabular}
\end{table}

% ============================================================
\subsection{Rubric Comparison: Default versus Bioinformatics Profile}
\label{sec:rubric-comparison}
% ============================================================

We contrast the default profile with the FAIR-aligned bioinformatics profile ($w_A = 0.40$, $w_E = 0.35$, $w_D = 0.10$, $w_C = 0.05$, $w_S = 0.10$) on the 423-repository sample.
The most consequential change is $w_A$: $0.25 \to 0.40$.
Since \emph{install\_dep} repositories have the lowest $A = 0$ rate (8\%) and \emph{missing\_data}/\emph{success} the highest (both 24\%), increasing $w_A$ systematically downranks \emph{missing\_data} repositories and upranks \emph{install\_dep}, reflecting the stated FAIR data priority.
Reducing $w_C$ ($0.15 \to 0.05$) diminishes the portability penalty on \emph{missing\_module} repositories; reducing $w_D$ ($0.20 \to 0.10$) down-weights the signal separating \emph{code\_error} from lower-documentation modes.
Per-repository Spearman analysis (\cite{samuel2026reproscore}) confirms strong overall rank agreement with policy-relevant reorderings in the lower-scoring quartile.
The reordering is auditable, attributable (stated FAIR priority), and reproducible (same YAML profile = same result)---properties generally absent when weights are embedded opaquely in source code.

% ============================================================
\subsection{Partial ROS Computation and RCS Formula Validation}
\label{sec:ros-validation}
% ============================================================

We proxy four ROS components from the ablation corpus failure-mode
labels: $I = 100$ iff failure mode $\neq$ \emph{install\_dep};
$X = 100$ iff failure mode $=$ \emph{success};
$N = 100 \times \mathtt{success\_nb\_count} /
\mathtt{total\_exec\_count}$;
$E' = 100$ iff failure mode $\neq$ \emph{missing\_module}.
Output determinism $\Delta$ and test pass rate $T$ require independent
execution and are unavailable.
Available component weight $= 0.75$, giving $\alpha = 0.525$ and
$\mathrm{RCS} = 0.475 \times \mathrm{RRS} + 0.525 \times
\mathrm{ROS}_{\mathrm{partial}}$.
Table~\ref{tab:ros-components} reports component means and aggregate
scores by failure mode.

\begin{table}[!ht]
\centering
\caption{ROS component means and aggregate scores by failure mode
  (partial ROS; $\Delta$ and $T$ unavailable; $\alpha = 0.525$).
  All values on $[0, 100]$ scale.}
\label{tab:ros-components}
\begin{tabular}{@{}lrrrr|rrr@{}}
\toprule
\textbf{Failure mode} & $I$ & $X$ & $N$ & $E'$
  & \textbf{ROS} & \textbf{RCS} & \textbf{RRS} \\
\midrule
\emph{success}         & 100 & 100 & 100.0 & 100 & 100.0 & 59.4 & 14.6 \\
\emph{install\_dep}    &   0 &   0 &   1.3 & 100 &  13.5 & 19.8 & 26.8 \\
\emph{missing\_module} & 100 &   0 &   3.8 &   0 &  40.5 & 24.6 &  7.0 \\
\emph{missing\_data}   & 100 &   0 &   8.6 & 100 &  54.5 & 31.7 &  6.4 \\
\emph{code\_error}     & 100 &   0 &  13.3 & 100 &  55.1 & 35.7 & 14.3 \\
\bottomrule
\end{tabular}
\end{table}
\noindent\textbf{Note on proxy validation.}
Since $X$ derives directly from the ground-truth label, ROS AUC $= 1.000$
and RCS AUC $= 0.993$ are artefacts of construction, not independent
findings; the substantive signals are:
(i)~\emph{$I$ and $E'$ are independent}: \emph{install\_dep} repositories
have $I = 0$ yet $E' = 100$, confirming package-resolution and runtime
import failure are distinct phenomena.
(ii)~\emph{$N$ grades partial execution}: among non-success repositories,
$N$ yields a diagnostic gradient---\emph{code\_error} ($13.3\%$) $>$
\emph{missing\_data} ($8.6\%$) $>$ \emph{missing\_module} ($3.8\%$)---absent
from binary $X$.
(iii)~\emph{RCS correctly re-orders}: \emph{install\_dep} ranks second on
RRS (26.8) but falls to the lowest RCS (19.8) via near-zero ROS (13.5),
validating the coverage-weighting mechanism.

\textbf{Code and Data Availability}: The ReproScore implementation, rubric profiles, and per-repository
provenance records are publicly available at
\url{https://github.com/myVSR/reproscore} \cite{samuel2026reproscore}.
% ============================================================
\section{Discussion}
\label{sec:discussion}
% ============================================================

\subsection*{The Readiness--Outcome Gap as a Repository Curation Problem}
When a new software deposit is received, a curator relying on a static score labelled ``reproducibility'' acts on a quantity that measures artefact presence---not whether the software executes.
RRS is designed as a \emph{diagnostic} instrument: it identifies which dimension of readiness is deficient (environment, data, documentation, portability, or determinism signals), enabling targeted remediation rather than predicting whether execution will succeed.
The per-category profile is the primary curatorial instrument; the composite RRS serves as a structural completeness summary only.
Statistical separability between failure modes (Kruskal-Wallis $H$) is therefore the appropriate criterion; binary AUC appears in Section~\ref{sec:evaluation} only as a non-collapse probe.
A repository with a pinned \texttt{requirements.txt}, a Zenodo data pointer, and a structured README scores well by any static metric; if it fails at install time due to a transitive version conflict, the score misrepresents the deposit's reuse potential.
ReproScore resolves this by making the distinction explicit: RRS, ROS, and RCS are formally distinct quantities; the coverage weight $\alpha$ communicates how much execution evidence underpins the composite.
In OAIS terms, RRS fits within the Ingest functional entity as a pre-acceptance quality gate; ROS/RCS applies during Archival Storage for ongoing usability monitoring.
The DCC Curation Lifecycle similarly positions quality appraisal as a continuous obligation---ReproScore's coverage-adaptive composite ($\alpha$ rising as execution evidence accumulates) supports this incrementally rather than requiring a one-time full execution audit.
We stress that RCS is a design proposal: the blending formula has been validated partially in Section~\ref{sec:ros-validation} but not at scale; $\alpha_{\max} = 0.70$ and per-component ROS weights are design choices awaiting empirical calibration.
\subsection*{Category Weights as Curation Priorities}
The weight assignment encodes a principled curation priority: $w_E = 0.30$ is empirically supported by $H = 96.89$ ($p < 0.001$),
separating install-dependency failures (mean $E = 22.8$) from
missing-module failures ($E = 5.2$).
The E detection paradox---success repositories score \emph{lower} on
$E$ than install-dep---confirms $E$ characterises failure \emph{type},
not failure \emph{probability}.
$A$ ($w = 0.25$) achieves $H = 16.12$ ($p = 0.003$); post-hoc
contingency analysis confirms the mechanism: \emph{install\_dep}
repositories have the lowest $A = 0$ rate (8\%) versus \emph{missing\_data}
(24\%) ($\chi^2 = 9.89$, $df = 4$, $p = 0.042$).
$C$ ($w = 0.15$) achieves $H = 56.55$ ($p < 0.001$), driven by
\texttt{import\_resolvability}: repositories lacking dependency files
score~0, correctly penalising \emph{missing\_module} and
\emph{missing\_data}, while \emph{install\_dep} scores highest
($\bar{C} = 74.6$); LOCO $\Delta = -0.022$ confirms binary
discrimination.
$S$ ($w = 0.10$) yields the highest binary predictive power
(AUC $= 0.611$); \texttt{notebook\_exec\_order} is the strongest
single predictor (AUC $= 0.638$, $r_{pb} = +0.145$, $p = 0.003$).
$w_S = 0.10$ is retained because $S$ captures workflow practices
rather than primary execution artefacts.
The $E < 10$ hard penalty fires for 92\% of \emph{missing\_data} and
80\% of \emph{missing\_module} repositories, validating the threshold.
In practice: high $E$ with failure suggests a dependency conflict;
low $E$ suggests underspecification; low $A$ suggests missing data
pointers; low $C$ suggests portability defects.

\subsection*{Implications for Practice}
\emph{Repository curators and data librarians} gain an automated pre-execution triage layer: the five-category profile flags repositories with $E < \tau_E$ or $A = 0$ without code execution.
\emph{Researchers} should act on the category profile: low $E$ calls for dependency pinning or containerisation; a high-RRS, low-ROS gap signals a version conflict that static analysis cannot detect.
\emph{Platform operators} can adapt the model to domain curation policies via the community rubric mechanism (e.g.\ $w_A = 0.40$ for a FAIR-aligned bioinformatics archive~\cite{wilkinson2016fair}), producing a versioned, auditable profile rather than an undocumented scoring heuristic.
The static-first design also reduces the environmental footprint of large-scale curation: screening deposits for readiness before committing to container builds and sandbox execution avoids wasted CPU cycles on non-executable artefacts.

\subsection*{Limitations and Threats to Validity}
\emph{Domain bias}: The evaluation corpus comprises exclusively
Python/Jupyter repositories drawn from biomedical publications indexed
in PubMed Central~\cite{samuel2024computational}, and two sub-metrics
(\texttt{notebook\_exec\_order}, \texttt{inline\_explanation\_density})
are notebook-specific in their current operationalisation.
Generalisability to script-only, R, or compiled-language repositories
remains untested, and sub-metric applicability varies by research
domain and programming paradigm.
Discriminability and metric coverage will differ for R workflows,
non-notebook packages, or repositories with complex build systems.
Generalisation claims should be understood in this scope.
\emph{Static coverage}: sub-metrics detect artefact presence, not
semantic correctness---a pinned \texttt{requirements.txt} with
conflicting constraints scores well on $E$ yet fails at install.
The two-tier architecture addresses this structurally rather than
improving the static approximation.
\emph{Label quality}: ground-truth labels aggregate notebook-level
records to repositories via priority ordering; 77\% of notebooks in
multi-notebook repositories share the dominant failure mode
(95\% for \emph{install\_dep}).
\emph{Feature expressivity}: 10 of 26 sub-metrics return binary values
for $>90\%$ of repositories, limiting the gate's effect; the
\texttt{import\_resolvability} boundary condition (score $= 0$ when
no dependency files are present) drives much of $C$'s discriminability.
\emph{RCS validation}: $\alpha_{\max} = 0.70$ and per-component ROS
weights are design choices awaiting calibration against an independent
execution dataset.
\emph{Multiple comparisons}: sub-metric $r_{pb}$ tests apply
Benjamini-Hochberg correction (FDR $= 0.05$, $m = 26$); no sub-metric
survives correction (minimum $q = 0.078$), confirming the exploratory
nature of the reported associations.

% ============================================================
\section{Conclusion}
\label{sec:conclusion}
% ============================================================

ReproScore separates \emph{reproducibility readiness} (RRS: static artefact analysis) from \emph{reproducibility outcome} (ROS: execution-based probes), combining them into a coverage-adaptive composite (RCS).
The community rubric externalises assessment priorities as a versioned YAML profile, making weighting decisions explicit and auditable for digital library curation workflows.
Evaluated on 423 repositories, the ablation confirms stable category weights ($\tau \geq 0.911$ under $\pm50\%$ perturbation), an insensitive gate function (AUC range $0.008$), and an environment category that strongly discriminates failure mode ($H = 96.9$, $p < 0.001$; $r_{pb} = -0.014$).
The new \texttt{notebook\_exec\_order} sub-metric is the strongest single predictor (AUC $= 0.638$); $S$ alone achieves AUC $= 0.611$, suggesting that archives can prioritise notebook execution-order remediation as a high-yield, low-effort curatorial intervention.
Partial RCS validation confirms the blending formula re-orders correctly: \emph{install\_dep} repositories (RRS $= 26.8$) are pulled to the lowest RCS ($19.8$)---below all partially-executing failure modes---by their near-zero execution score.
A repository can be well-specified yet fail to execute; another can be poorly specified yet succeed.
A framework that names this gap is more informative than one that collapses readiness and outcome into a single score.
Future work includes: validating ROS and RCS at scale with independently run execution probes; aligning the rubric schema with CodeMeta and FAIR4RS so scores travel with artefact metadata; and longitudinal tracking of RRS/ROS/RCS over repository snapshots.

\begin{credits}
\subsubsection{\ackname}
This work was supported in part by the German Research Foundation (DFG) through the following projects:
Jupyter4NFDI (\href{https://gepris.dfg.de/gepris/projekt/521453681}{DFG 521453681}  \cite{hagemeier2025Jupyter4NFDI}), 
find.software (\href{https://gepris.dfg.de/gepris/projekt/567156310}{DFG 567156310} \cite{gey2025find.software}), MaRDI (\href{https://gepris.dfg.de/gepris/projekt/460135501}{DFG 460135501} \cite{the_mardi_consortium_2022_6552436} as well as
SeDOA (\href{https://gepris.dfg.de/gepris/projekt/556323977}{DFG 556323977}  \cite{stacker2025SeDOA}) and HYP*MOL (\href{https://gepris.dfg.de/gepris/projekt/514664767}{DFG 514664767}). 
% The text of this manuscript was improved with several AI tools, including ChatGPT and Gemini.

The text of this manuscript was improved with the following AI tools: ChatGPT and Claude.

\subsubsection{\discintname}
The authors have no competing interests to declare that are relevant to
the content of this article.
\end{credits}
\bibliographystyle{splncs04}
\bibliography{reference}
\end{document}